\documentclass[12pt]{amsart}
\usepackage{amsfonts,amsthm}
\usepackage{amssymb,amsmath}
\usepackage{amsfonts,mathrsfs}
\usepackage{xcolor}

\begin{document}

\noindent{\bf Felix Klein's
``\"Uber die Differentialgesetze f\"ur die Erhaltung von Impuls und Energie in der Einsteinschen  Gravitationstheorie'': an English translation}

\bigskip
\centerline{Chiang-Mei Chen$^{1,2}$, James M. Nester$^{1,3,4}$ and Walter Vogel$^5$}

\medskip

\noindent $^1$ Department of Physics, National Central University, Chungli 32001, Taiwan

\noindent $^2$ Center for High Energy and High Field Physics (CHiP), National Central University,
Chungli 32001, Taiwan

\noindent $^3$ Graduate Institute of Astronomy, National Central University, Chungli 32001, Taiwan

\noindent $^4$ Leung Center for Cosmology and Particle Astrophysics, National Taiwan University, Taipei 10617, Taiwan

\noindent $^5$ Department of Chemistry, National Central University, Chungli, 32001, Taiwan,

\medskip
\noindent email: cmchen@phy.ncu.edu.tw, nester@phy.ncu.edu.tw, vogelw@ncu.edu.tw

\bigskip

\noindent{\textbf{Abstract}}

\emph{We present an English translation of a second 1918 paper by Felix Klein which follows up on his earlier work.}

\section{Translator's Preface}

In 1918, following up on his paper~\cite{KleinI} about David Hilbert's Foundation's of Physics I~\cite{Hilbert} (which we have recently translated), Felix Klein published another work~\cite{KleinII} about the law of conservation of energy and momentum in Einstein's theory of gravitation, general relativity. This work, which has for too long been neglected, includes some interesting analysis regarding the gravitational energy-momentum  expressions of Einstein,  Hilbert, Lorentz and Weyl.  The topic of gravitational energy-momentum and its localization had been at that time---and, notwithstanding considerable progress, still remains a century later---an unsettled issue.  For the detailed story concerning the related exchanges between Einstein, Hilbert and Klein and the inception of Noether's theorems see Refs.~\cite{Brading05, KS, Rowe99}.

Some years ago the senior member of our team (JMN) began, relying on his long unused undergraduate German and Google Translate, to make a translation of Klein's papers, which we believe include some long forgotten insights. We are fortunate to have recently acquired the help of a native German speaker (WV) to refine our effort into a presentable form. We feel that our translation has now finally reached a form where it can be useful to others, and so want to share it with anyone who may be interested.

The page by page layout, the equation numbers, and footnotes in this version of our translation are from the paper as it appears in Vol.~1 of Klein's collected works~\cite{KleinII}---so anyone who cares to can easily compare our translation with the original. We chose to follow the Klein collected works version, as it includes some additional footnotes that do not appear in the journal version; readers may find the remarks regarding Emmy Noether especially interesting. Our translation is a work in progress. We welcome corrections and comments on the translation and on any errors.

\newpage

\textbf{XXXII. On the differential laws for the conservation of momentum and energy in the Einstein theory of gravitation.}\\

\footnotesize{[News of the Kgl. Society of Sciences at G\"ottingen. Mathematical-Physical class. (1918.) presented at the sitting of
19 July 1918.\footnote{The manuscript did not take its final form until the middle of September this year.}]}
\bigskip

---------------
\bigskip

In the continuation of the investigations which I submitted to the Society of Sciences on 25 January of this year%
\footnote{See the conclusion of the year 1917 of these news: ``On Hilbert's First Note on the Foundations of Physics.''}, I have succeeded in describing the forms of the differential laws for the conservation of momentum and energy by various authors for Einstein's gravitation theory,%
\footnote{Primarily to be considered here is Einstein's comprehensive essay of 1916, ``The Foundations of the General Relativity Theory" (Leipzig), and the Communication to the Berlin Academy, ``Hamilton's Principle and General Relativity Theory" (Session Report, 26 October 1916), by Hilbert the already mentioned note (G\"ottingen news of November 20, 1915), of Lorentz the four articles published on the occasion of a lecture held in Leiden by the Amsterdam Academy from March to June 1916 --- ``Einstein's theory of gravity''--- see in particular Article III of April and September 1916 and Article IV of October 1916 and May 1917, respectively. I also mention the recently published book of Weyl ``Space - Time - Matter'' (Berlin, 1918), which I continue to refer to. [Weyl's book is already in its third edition; The first edition is always quoted in the text.]}
deduced from a uniform point of view and, if I am not mistaken, a much improved insight into their meaning and reciprocal relationships. As one shall see, I have in the following description actually no longer to calculate, but only to make use of the most elementary formulas of the classical variational calculus.

\newpage

XXXII. Differential Form of Conservation Laws in Gravitational Theory. 569

\bigskip

For the sake of brevity, I shall here, as well as for the nomenclature, refer to my previous note. As an actual reason for the progress I have now made, is that the infinitesimal transformation
\begin{equation}%(1)
\delta w^\tau = p^\tau
\end{equation}
considered at that time is no longer subject to the restriction, at the boundary of the region of integration in a suitable manner (namely, the first and second differential quotient $p^\tau_\varrho$, $p^\tau_{\varrho\sigma}$) to vanish. This results in boundary components for the relevant integrals, a closer examination of which provides all the rest. For the particular purpose envisaged here, it suffices to consider only the first of the two integrals considered earlier:
\begin{equation}%(2)
I_1 = \int\!\int\!\int\!\int K d\omega.
\end{equation}
From practical reasons the consideration is further divided so that $K$ depends arbitrarily on the functions $g^{\mu\nu}$, $g^{\mu\nu}_{\varrho}$, $g^{\mu\nu}_{\varrho\sigma}$, and then as an invariant (which is not yet determined) against arbitrary transformations of the world parameters $w$, and finally as an invariant of a certain type.

By applying (1) to (2) in such a way, a series of differential relations are produced, which are identically satisfied by $K$. Now I turn to physics, not limiting myself, as in the previous case, to the case of the free electromagnetic field, but immediately  presupposing an arbitrary ``material'' field. If one combines Hilbert's approach with that of Einstein, the corresponding ten field equations of gravitation are known to be written in the simple form%
\footnote{See, e.g., B. Herglotz in the Saxon Reports, 1916, p. 202, formula (16). --- For the sake of accuracy, I also note that the constant $\chi$ (which in my previous note to Hilbert I denoted as $\alpha$) has the value
$$ \chi = 1,87 \cdot 10^{-27} \cdot {\rm cm}^{+1} {\rm gr}^{-1} $$
with the underlying $ds^2$ having the dimension and the sign which agrees with the $d\tau^2$ of the special theory of relativity:
$$ d\tau^2 = dt^2 - \frac{dx^2 + dy^2 + dz^2}{c^2} \sim {\rm sec}^2. $$}:
\begin{equation}%(3)
K_{\mu\nu} - \chi T_{\mu\nu} = 0,\footnote{[In reprinting, the signs of $T_{\mu\nu}$, $\mathfrak{T}_{\mu\nu}$, $T^\sigma_\tau$, $\mathfrak{T}^\sigma_\tau$, $t^\sigma_\tau$, $\mathfrak{t}^\sigma_\tau$ have been reversed here and in the following in order to be in harmony with the usual names in physics. For example, $T_{44}$ is then positive. K.]}
\end{equation}%
$K_{\mu\nu}$ is the Lagrangian derivation with respect to the $g^{\mu\nu}$ associated with $I_1$, which is divided by $\sqrt{g}$,  $T_{\mu\nu}$ are the energy components of the matter. \textit{The transition to the various forms of the conservation theorems is simply based on the principle that $\chi T_{\mu\nu}$ is inserted for $K_{\mu\nu}$ in the identities derived for $K$.}

\newpage

570 On the Erlangen Program.
\bigskip

\centerline{\S 1.}
\medskip

\centerline{\textbf{Infinitesimal transformation of the $g^{\mu\nu}$.}}
\bigskip

In order to give the reader all the means of control in hand, here I describe the small intermediate calculation which determines the $\delta g^{\mu\nu}$ corresponding to the infinitesimal transformation (1) of the $w$.

Instead of (1), I first write
$$ \bar w^\mu = w^\mu + p^\mu $$
(where for the ``helping vector'' $p$ and its differential quotients $p_\varrho$, $p_{\varrho\sigma}$ it will further be expected that all the higher-order terms can be neglected against the linear ones). We have then
$$
d\bar w^\mu = dw^\mu + \sum p^\mu_\tau dw^\tau.
$$
Now, according to their definition, the $g^{\mu\nu}$ are cogredient to the products $dw^\mu dw^\nu$. Therefore:
$$
\bar g^{\mu\nu}(\bar w) = g^{\mu\nu}(w) + \sum g^{\mu\tau}(w) p^\nu_\tau + \sum g^{\nu\tau}(w) p^\mu_\tau.
$$
But
$$
g^{\mu\nu}(\bar w) = \bar g^{\mu\nu}(w) + \sum g^{\mu\nu}_\tau(w) p^\tau.
$$
\textit{Now it is the difference $g^{\mu\nu}(w) - \bar g^{\mu\nu}(w)$ which I have named in my previous note $\delta g^{\mu\nu}$, and which I shall now denote by $p^{\mu\nu}$ in accordance with Hilbert's note.} Then:
\begin{equation}%(4)
\delta g^{\mu\nu} = p^{\mu\nu} = \sum_\tau (g^{\mu\nu}_\tau p^\tau - g^{\mu\tau} p^\nu_\tau - g^{\nu\tau} p^\mu_\tau).
\end{equation}
The differential quotients of the $p^{\mu\nu}$ with respect to the $w$ are, as in Hilbert, designated by $p^{\mu\nu}_{\varrho}$, $p^{\mu\nu}_{\varrho\sigma}$.

I also note the value which $p^{\mu\nu}$ obtains in the case of constant $p^\tau$ which will later be considered (where I write $\underset{0}{p}{}^{\mu\nu}$ and $\underset{0}{p}{}^\tau$, respectively):
\begin{equation}%(5)
\underset{0}{p}{}^{\mu\nu} = \sum_\tau g^{\mu\nu}_\tau \underset{0}{p}{}^\tau.
\end{equation}
In this case it is as if the $g^{\mu\nu}$ were fixed functions of the $w$ [scalars] (not a substitution induced by the respective transformation of the $w$).

\bigskip

\centerline{\S 2.}
\medskip

\begin{center}
\textbf{Calculating $\delta I_1$ under the sole assumption that $K$ is a \\ function of $g^{\mu\nu}$, $g^{\mu\nu}_\varrho$, $g^{\mu\nu}_{\varrho\sigma}$. --- The fundamental theorem}.
\end{center}
%%%jmn alternatives: ``principle statement'', ``main theorem'', ``principle theorem'', ``fundamental theorem''

\bigskip

This means that $K$ does not depend explicitly on the $w$. We then have for our infinitesimal transformation [which extends to the dependent as well as independent variables of the integral $I_1$] first:

\newpage

XXXII. Differential Form of Conservation Laws in Gravitational Theory. 571
\bigskip

\begin{eqnarray}%(6)
\delta I_1 &\!=\!& - \int\!\!\int\!\!\int\!\!\int \sum_{\mu\nu} \left( \frac{\partial \sqrt{g} K}{\partial g^{\mu\nu}} p^{\mu\nu} \!+\! \sum_\varrho \frac{\partial \sqrt{g} K}{\partial g^{\mu\nu}_\varrho} p^{\mu\nu}_\varrho \!+\! \sum_{\varrho,\sigma} \frac{\partial \sqrt{g} K}{\partial g^{\mu\nu}_{\varrho\sigma}} p^{\mu\nu}_{\varrho\sigma} \right) dS
\\ \nonumber
&& + \int\!\!\int\!\!\int \sqrt{g} K ( p^I dw^{II} dw^{III} dw^{IV} + \dots + p^{IV} dw^I dw^{II} dw^{III} );
\end{eqnarray}
$dS$ is written for $dw^I dw^{II} dw^{III} dw^{IV}$, the triple integral in the known vectorial fashion is extended over the edge of the integration domain of $I_1$.%
\footnote{[There is an essential difference, and at the same time progress, compared to my previous note in that I do not presuppose anything about the behavior of the $p^\tau$, $p^{\mu\nu}$, $p^{\mu\nu}_\varrho$, $p^{\mu\nu}_{\varrho\sigma}$ at the edge of the region of integration. K.]}

Here, we shall eliminate the differential quotients $p^{\mu\nu}_\varrho$, $p^{\mu\nu}_{\varrho\sigma}$ occurring under the quadruple integral uniquely according to the old method of Lagrange, by means of two fold partial integration, then replace $p^{\mu\nu}$ by its value (4), and eliminate the differential quotients $p^\nu_\tau$ and $p^\mu_\tau$ which result thereby in a fourth partial integration. \textit{We find thus:
\begin{eqnarray}%(7)
\delta I_1 &=& - \int\!\!\int\!\!\int\!\!\int \sum_\tau \left( \sqrt{g} \sum_{\mu\nu} K_{\mu\nu} g^{\mu\nu}_\tau + 2 \sum_\sigma \frac{\partial \sqrt{g} K^\sigma_\tau}{\partial w^\sigma} \right) p^\tau \cdot dS
\\ \nonumber
&& + \int\!\!\int\!\!\int \sqrt{g} (\varepsilon^I dw^{II} dw^{III} dw^{IV} + \dots + \varepsilon^{IV} dw^I dw^{II} dw^{III}),
\end{eqnarray}
where the following abbreviations are introduced:}

1. $K_{\mu\nu}$ is the Lagrangian derivative divided by $\sqrt{g}$, which was already used in (3):
\begin{equation}%(8)
K_{\mu\nu} = \left( \frac{\partial \sqrt{g} K}{\partial g^{\mu\nu}} - \sum_\varrho \frac{\partial \left( \frac{\partial \sqrt{g} K}{\partial g^{\mu\nu}_\varrho} \right)}{\partial w^\varrho} + \sum_{\varrho\sigma} \frac{\partial^2 \left( \frac{\partial\sqrt{g} K}{\partial g^{\mu\nu}_{\varrho\sigma}} \right)}{\partial w^\varrho \partial w^\sigma} \right):\sqrt{g};
\end{equation}

2. $K^\sigma_\tau$ is the following linear combination of the $K_{\mu\nu}$:
\begin{equation}%(9)
K^\sigma_\tau = \sum_\mu K_{\mu\tau} g^{\mu\sigma};
\end{equation}

3. $\varepsilon^\sigma$, for $\sigma = 1, 2, 3, 4$, is a five-term expression, which I write in advance (by referring especially to the term resulting from the fourth partial integration):
\begin{equation}%(10)
\varepsilon^\sigma = \eta^\sigma + 2 \sum_\tau K^\sigma_\tau p^\tau.
\end{equation}

Here is then

4. $\eta^\sigma$ is a four-term expression:
\begin{equation}%(11)
\eta^\sigma = K p^\sigma - \sum_{\mu\nu} \frac{\partial K}{\partial g^{\mu\nu}_\sigma} p^{\mu\nu} - \sum_{\mu\nu\varrho} \frac{\partial K}{\partial g^{\mu\nu}_{\varrho\sigma}} p^{\mu\nu}_\varrho + \frac1{\sqrt{g}} \sum_{\mu\nu\varrho} \frac{\partial \left(\frac{\partial \sqrt{g} K}{\partial g^{\mu\nu}_{\varrho\sigma}} \right)}{\partial w^\varrho} p^{\mu\nu}.
\end{equation}

\newpage

572 On the Erlangen Program.
\bigskip

The fourfold integral which is presented in the new expression (7) of $\delta I_1$ is henceforth, after omitting the minus sign, to be called \textit{the integral A}.

In addition, we will transform the three-fold integral occurring in (7) into a second quadruple integral by the elementary formation of a divergence,
\begin{equation}%(12)
\int\!\!\int\!\!\int\!\!\int \left( \frac{\partial \sqrt{g} \varepsilon^I}{\partial w^I} + \dots + \frac{\partial \sqrt{g} \varepsilon^{IV}}{\partial w^{IV}} \right) dS,
\end{equation}
which shall be called \textit{integral B}. Thus
\begin{equation}%(13)
\delta I_1 = - A + B.
\end{equation}

\textit{Here we have the important observation that $A \equiv B$ if we choose the $p^\tau$ constant, i.e. = $\underset{0}{p}{}^\tau$}.

In fact, the original value (6) of $\delta I_1$ vanishes identically, \textit{because $K$ does not explicitly contain the $w$}, using the values of the $\underset{0}{p}{}^{\mu\nu}$ given in (5).

From $A \equiv B$, however, we conclude with full arbitrariness regarding the choice of the integration domain, that the integrands of $A$ and $B$ must also agree. We have
\begin{eqnarray}%(14)
\sum_\tau \left( \sqrt{g} \sum_{\mu\nu} K_{\mu\nu} g^{\mu\nu}_\tau + 2 \sum_\sigma \frac{\partial \sqrt{g} K^\sigma_\tau}{\partial w^\sigma} \right) \underset{0}{p}{}^\tau \equiv \sum_\sigma \frac{\partial \sqrt{g} \, \underset{0}{\varepsilon}{}^\sigma}{\partial w^\sigma}
\\ \nonumber
\equiv \sum_\sigma \frac{\partial \sqrt{g} \, \underset{0}{\eta}{}^\sigma}{\partial w^\sigma} + 2 \sum_{\sigma\tau} \frac{\partial \sqrt{g} K^\sigma_\tau}{\partial w^\sigma} \underset{0}{p}{}^\tau.
\end{eqnarray}
\textit{This identity will henceforth be called the fundamental theorem}.
%%%jmn  ``fundamental theorem'' or ``main theorem'' or ``main statement'' or ``principle statement''

We can of course remove the terms with $K^\sigma_\tau$ on both sides. Let us write $\underset{0}{\eta}{}^\sigma$ as follows as a function of the $\underset{0}{p}{}^\tau$:
\begin{equation}%(15)
\underset{0}{\eta}{}^\sigma = 2 \sum U^\sigma_\tau \underset{0}{p}{}^\tau
\end{equation}
(where we add 2 on the right hand side because it is later indicated to divide by a 2). Here (using the usual designation $\delta^\sigma_\tau$ for 1 or 0, depending on $\sigma = \tau$ or $\sigma \ne \tau$):
\begin{equation}%(16)
2 U^\sigma_\tau = K \delta^\sigma_\tau - \sum_{\mu\nu} \frac{\partial K}{\partial g^{\mu\nu}_\sigma} g^{\mu\nu}_\tau - \sum_{\mu\nu\varrho} \frac{\partial K}{\partial g^{\mu\nu}_{\varrho\sigma}} g^{\mu\nu}_{\varrho\tau} + \frac1{\sqrt{g}} \sum_{\mu\nu\varrho} \frac{\partial \left( \frac{\partial\sqrt{g} K}{\partial g^{\mu\nu}_{\varrho\sigma}} \right)}{\partial w^\varrho} g^{\mu\nu}_\tau.
\end{equation}
\textit{The fundamental theorem now takes the following form}:
%%% or ``main theorem'', ``fundamental theorem'', ``principle theorem''
%
\begin{equation}%(17)
\sum_{\mu\nu} \sqrt{g} K_{\mu\nu} g^{\mu\nu}_\tau \equiv 2 \sum_\sigma \frac{\partial \sqrt{g} U^\sigma_\tau}{\partial w^\sigma}, \quad \hbox{for} \; \tau = 1, 2, 3, 4.
\end{equation}
The expressions on the left are thus transformed into elementary divergences.

\newpage

XXXII. Differential Form of Conservation Laws in Gravitational Theory. 573
\bigskip

\centerline{\S 3.}
\medskip

\begin{center}
\textbf{Simplified description of the formulas. --- An extension of \\ the fundamental theorem}.
\end{center}
%%% alternatives: ``main theorem'', ``principle theorem'', ``fundamental theorem''.

\bigskip

In the preceding paragraph, I chose the representation as seems appropriate for their later invariant-theoretic evaluation and, by the way, confirms with old habits. In the meantime many things can be abbreviated according to Einstein's proposals:

1. Much can be saved by replacing the product of $\sqrt{g}$ with a quantity designated by a large Latin letter by the corresponding German letter. Thus, $\sqrt{g} K$ by $\frak{K}$, $\sqrt{g} K_{\mu\nu}$ by $\frak{K}_{\mu\nu}$, $\sqrt{g} K^\sigma_\tau$ by $\frak{K}^\sigma_\tau$, $\sqrt{g} U^\sigma_\tau$ by $\frak{U}^\sigma_\tau$. (In the sense of this agreement it will be possible to write an \textit{elementary divergence} as follows:
\begin{equation}%(18)
\frac{\partial \frak{W}^I}{\partial w^I} + \frac{\partial \frak{W}^{II}}{\partial w^{II}} + \frac{\partial \frak{W}^{III}}{\partial w^{III}} + \frac{\partial\frak{W}^{IV}}{\partial w^{IV}} = \frak{Div}.
\end{equation}
Here the $\frak{W}^I$, \dots, $\frak{W}^{IV}$ should depend only on the $g^{\mu\nu}$, $g^{\mu\nu}_\varrho$, so that our $\frak{Div}$ is a special case of the functions $\frak{K}$ so far considered.)

2. Furthermore, in summation expressions the summation letters can be omitted by noting that the indices are always summed twice (once up and once down).

3. Finally, the sums can be omitted on the same grounds.

We shall make more or less use of the abbreviation as soon as it is fit for us. The formulas (17) for example, are rewritten as follows:
\begin{equation}%(19)
\frak{K}_{\mu\nu} g^{\mu\nu}_\tau \equiv 2 \frac{\partial \frak{U}^\sigma_\tau}{\partial w^\sigma}.
\end{equation}

In connection with this, I shall now consider a remarkable generalization of the formula (17), or (19).

For the divergences introduced ($\frak{Div}$), the Lagrange derivatives will vanish identically in a known manner:
\begin{equation}%(20)
\frak{Div}_{\mu\nu} = 0.
\end{equation}
If, therefore, in (19) instead of $\frak{K}_{\mu\nu}$ we substitute the Lagrange derivatives of a function $\frak{K}^*$, which is related to $\frak{K}$ by an equation:
\begin{equation}%(21)
\frak{K}^* = \frak{K} + \frak{Div},
\end{equation}
the left-hand side of (19) remains unchanged, however on the right side instead of $\frak{U}^\sigma_\tau$ a new function $\frak{U}^*{}^\sigma_\tau$ occurs. We then have

\newpage

574 On the Erlangen Program.

\bigskip

\begin{equation}%(22)
\frak{K}_{\mu\nu} g^{\mu\nu}_\tau = 2 \frac{\partial \frak{U}^*{}^{\sigma}_\tau}{\partial w^\sigma}.
\end{equation}

\textit{Thus the formula (19) is remarkably generalized}. (Of course, the $\frak{U}^*{}^{\sigma}_\tau$ of the $\frak{U}^\sigma_\tau$, if the $\tau$ is fixed, differs only by terms whose elementary divergences $\sum\frac{\partial}{\partial w^\sigma}$ vanish identically.)

\bigskip

\centerline{\S 4.}
\medskip

\centerline{\textbf{Invariant theoretical viewpoints.}}
\bigskip

We shall now assume, in the sense of the general theory of relativity, that $K$ is invariant under the group of all the transformations of the $w$ (which, of course, we must think of as ``extended'' by assuming the corresponding transformations of the $g^{\mu\nu}$).

Since $d\omega$ is an invariant naturally, the same is true of the integral $I_1$.

$K_{\mu\nu}$ appears as a contra-gredient tensor; the complex of the 16 quantities $K^\sigma_\tau$ as a mixed tensor.

Furthermore (by thinking of the helping-vector $p$ as transformed as the $dw$), we may denote the $\varepsilon^\sigma$, $\eta^\sigma$ as co-gredient vectors.%
\footnote{[That $\varepsilon^\sigma$ and $\eta^\sigma$ are co-gredient vectors is most easily seen by comparing their expressions
(Eq.~(10) and (11)) with the formulas (8), (9) and (14) of the first communication from Hilbert on the Foundations of physics, (loc.~cit.). K.]}

If we now write $A$ as follows:
\begin{equation}%(23)
A = \int\!\!\int\!\!\int\!\!\int \sum\left( \frac{\sqrt{g} \sum K_{\mu\nu} g^{\mu\nu}_\tau + 2 \sum \frac{\partial \sqrt{g} K^\sigma_\tau}{\partial w^\sigma}}{\sqrt{g}} \cdot p^\tau \right) d\omega,
\end{equation}
the system of magnitudes multiplied by the different $p^\tau$ appears as a contra-gredient vector (it is, in the sense of my earlier note, the ``vectorial divergence'' of the tensor $K_{\mu\nu}$).

Correspondingly, we obtain an invariant from $B$:
\begin{equation}%(24)
\frac1{\sqrt{g}} \sum \frac{\partial\sqrt{g} \varepsilon^\sigma}{\partial w^\sigma};
\end{equation}
we shall call it (again in the sense of my previous note) the ``scalar divergence'' of the vector $\varepsilon$ (formed with the help of the vector $p$).

Equally, the two components of (24):
\begin{equation}%(25)
\frac1{\sqrt{g}} \sum \frac{\partial \sqrt{g} \eta^\sigma}{\partial w^\sigma} \quad \hbox{and} \quad \frac1{\sqrt{g}} \sum \frac{\partial (\sqrt{g} K^\sigma_\tau p^\tau)}{\partial w^\sigma}
\end{equation}
will in themselves be invariants.

\newpage

XXXII. Differential Form of Conservation Laws in Gravitational Theory. 575

\bigskip

\textit{But how is it with the $U^\sigma_\tau$ which were derived under the assumption of constant $p^\tau$ ($= \underset{0}{p}{}^\sigma$)?}

Constant $p^\tau$ no longer remains constant in the case of arbitrary transformations of the $w$, but only in the case of the ``affine'' transformations:
$$
\bar w^\varrho = a^\varrho_1 \cdot w^I + \dots + a^\varrho_4 \cdot w^{IV} + c^\varrho.
$$
Of course one should think of the $g^{\mu\nu}$ as transformed correspondingly (i.e. linearly with constant coefficients).
Keep in mind that the individual $g^{\mu\nu}$ are functions of $w^\varrho$.

We may then say:

\textit{$U^\sigma_\tau$ is a mixed tensor of the thus expanded affine group}.

This does not preclude that, according to our equations (14), (17), and the notation (13), the expression independent of the $p^\tau$
\begin{equation}%(26)
\frac2{\sqrt{g}} \sum \frac{\partial (\sqrt{g} (U^\sigma_\tau + K^\sigma_\tau))}{\partial w^\sigma}
\end{equation}
is a contra-gredient vector of the \textit{general} group.

This is a very curious circumstance, which is fundamental to the later discussion.

If, according to (21), $\frak{K}$ is replaced by any $\frak{K}^*$, and suppose the assumption that the $\frak{W}^I$ \dots $\frak{W}^{IV}$ occurring in (18) are equal to the components multiplied by $\sqrt{g}$ of a vector $W^I$ \dots $W^{IV}$ of the affine group:
\begin{equation}%(27)
\frak{W}^\sigma = \sqrt{g} W^\sigma,
\end{equation}
then exactly the same fact as in (26) is given for the more general expressions,
\begin{equation}%(28)
\frac2{\sqrt{g}} \sum \frac{\partial (\sqrt{g} (U_\tau^{*\sigma} + K^\sigma_\tau))}{\partial w^\sigma}.
\end{equation}
\bigskip

\centerline{\S 5.}
\medskip

\centerline{\textbf{Identities for which our $K$ is an invariant of the general group}.}
\bigskip

We now pursue the idea that: \textit{since $K$ is an invariant of our general group, it follows, in the case of arbitrary values of the $p^\tau$}:
\begin{equation}%(29)
\delta I_1 = \delta \int\!\!\int\!\!\int\!\!\int K d\omega = 0.
\end{equation}
(Conversely, if the relation (29) holds for any $p^\tau$, then $I_1$ and hence $K$ will be an invariant of the general group, since all finite transformations of the $w^\tau$ are composed of the infinitesimal $\delta w^\tau = p^\tau$).

\newpage

576 On the Erlangen Program.
\bigskip

We thus obtain from the formulas of \S\S 2 and 3 a large number of differential relations,  which the invariant $K$ (which is not yet individualized at all) has to obey identically.

1. We take, as in my previous note, the $p^\tau$, without otherwise restricting their arbitrariness, that the vector $\varepsilon^\sigma$ and thus the associated boundary integral simply vanishes. This obviously implies that $p^\tau$, $p^{\mu\nu}$, and $p^{\mu\nu}_\varrho$ vanish along the boundary, i.e. the nullity of $p^\tau$, $p^\tau_\varrho$, $p^\tau_{\varrho\sigma}$. Then, according to (13), $A = 0$, i.e. in any region of integration and any assumption of the $p^\tau$ in the interior of the region. We conclude according to (23) \textit{that the vectorial divergence of the tensor $K_{\mu\nu}$ must be identical to zero}. In formulas:
\begin{equation}%(30)
\frac{\sqrt{g} \sum K_{\mu\nu} g^{\mu\nu}_\tau + 2 \sum \frac{\partial \sqrt{g} K^\sigma_\tau}{\partial w^\sigma}} {\sqrt{g}} \equiv 0 \qquad (\hbox{for}\; \tau = 1, 2, 3, 4).
\end{equation}
These are the identities (12) of my previous note, which I shall now call \textit{the identities A}. --- It is clear: since it is a vector, the left-hand sides of (30), set up for any coordinate system, will be equal to well-known linear combinations of their original values. The vanishing of the transformed expressions thus means nothing but the vanishing of the original expressions.

2. As a consequence of the identities (30), the integral $A$ now drops away at \textit{any} $p^\tau$. Thus, according to (13), (29) the integral $B$ also always vanishes. Again, we consider that the integration domain and the vector $p^\tau$ can be assumed quite arbitrarily. It follows that the integrand of $B$, i.e. the scalar divergence of the vector $\varepsilon$, must be identically zero:
\begin{equation}%(31)
\frac1{\sqrt{g}} \sum \frac{\partial \sqrt{g} \varepsilon^\sigma}{\partial w^\sigma} \equiv 0,
\end{equation}
or what is the same:
$$%(31')
(31') \hspace{4cm}
\frac1{\sqrt{g}} \sum \frac{\partial \sqrt{g} (\eta^\sigma + 2 \sum K^\sigma_\tau p^\tau)}{\partial w^\sigma} \equiv 0. \hspace{4cm}
$$

In this one formula (31) and (31$'$) there are still a great many individual equations with the arbitrariness of the $p^\tau$. Consider the terms which arise from $\sum K^\sigma_\tau p^\tau$ in the differentiation, and consider that $\eta^\sigma$ is composed of terms which contain $p^{\mu\nu}$, $p^{\mu\nu}_\varrho$ linearly, while the $p^{\mu\nu}$ itself is linear in the $p$ and its differential quotients with $w$. But the $\eta^\sigma$ in (31) as in~(31$'$) again are differentiated with respect to  $w$. We conclude that the left-hand sides of (31) and (31$'$) are homogeneously linear in the $p^\tau$ and their first, second, and third differential quotients.

\newpage

\noindent XXXII, Differential Form of Conservation Laws in Gravitational Theory. 577

\bigskip

\noindent
Since these can all be assumed independently of one another, we have on the whole
$$ 4 \left( 1 + 4 + \frac{4 \cdot 5}{1 \cdot 2} + \frac{4 \cdot 5 \cdot 6}{1 \cdot 2 \cdot 3} \right) = 140 $$
equations. I will call these \textit{identities $B$}.

It is worthwhile at least schematically to account for these 140 equations. I will not be in contradiction with the name introduced in (15), when I write:
\begin{equation}%(32)
\eta^\sigma = 2 \left( \sum U^\sigma_\tau p^\tau + \sum U^{\sigma, \sigma'}_\tau p^\tau_{\sigma'} + \sum U^{\sigma,\sigma'\sigma''}_\tau p^\tau_{\sigma'\sigma''} \right)
\end{equation}
(each to be summed over all the twice repeated indices that occur, and indices that are separated by no comma are commutable, not the comma-separated ones. Then there are 16 (1 + 4 + 10) = 240 quantities $U$). --- The equation (31$'$) will now by omitting the preceding factor $\frac2{\sqrt{g}}$ and again writing $\frak{U}$ for $\sqrt{g} U$ become as follows:

1. 4 Equations corresponding to the terms with $p^\tau$:
\begin{equation}%(33)
\sum (\frak{U}^\sigma_{\tau,\sigma} + \frak{K}^\sigma_{\tau,\sigma}) \equiv 0,
\end{equation}

2. 16 Equations corresponding to the terms with $p^\tau_\sigma$:
\begin{equation}%(34)
\frak{U}^\sigma_\tau + \frak{K}^\sigma_\tau + \sum_{\sigma'} \frak{U}^{\sigma',\sigma}_{\tau,\sigma'} \equiv 0,
\end{equation}

3. 40 Equations corresponding to the terms with $p^\tau_{\sigma' \sigma''}$:
\begin{equation}%(35)
\frak{U}^{\sigma',\sigma''}_\tau + \frak{U}^{\sigma'',\sigma'}_\tau + \sum_{\sigma} \frak{U}^{\sigma,\sigma'\sigma''}_{\tau,\sigma} \equiv 0,
\end{equation}
[Editor's note: Here the last term should have a coefficient of 2.]

4. 80 Equations which correspond to the terms with $p^\tau_{\sigma{\sigma'}{\sigma''}}$:
\begin{equation}%(36)
\frak{U}^{\sigma,\sigma'\sigma''}_\tau + \frak{U}^{\sigma',\sigma''\sigma}_\tau + \frak{U}^{\sigma'',\sigma\sigma'}_\tau \equiv 0.
\end{equation}
\textit{I have not investigated the dependencies that may exist between these 140 equations}.

Moreover, the following conclusions are immediately drawn:

a) The identities $A$ (= (30)) and $B$ (= (33), (34), (35), (36)) together form the sufficient conditions that a function $K$ of the $g^{\mu\nu}$, $g^{\mu\nu}_\varrho$, $g^{\mu\nu}_{\varrho\sigma}$ is an invariant of our general group.

b) But the left-hand sides of (33), multiplied by $\frac{2}{\sqrt{g}}$, are directly identical with the left-hand sides of (30) because of the fundamental theorem of \S2.
%%% ``principle theorem'', ``main theorem'', ``fundamental theorem''

c) Thus, the $B$ alone are the sufficient conditions for the invariance of $K$.

d) But the $A$ alone are not. For the equations $A$ will also exist if $K$ is replaced by $K^* = K + {\rm Div}$, more generally, if $K$ is such a function which increases by a divergence with any transformation of the $w$.

\bigskip

{\tiny{Klein, Collected math. Treatises. I. 37}}

\newpage

578 On the Erlangen Program.

\bigskip

e) Therefore the identities $B$ cannot be deduced generally from the $A$.

For us, however, with respect to the physical conclusions to be developed, the $A$ is in the first rank. The following are the three forms which they can assume according to the preceding:
\begin{eqnarray}%(37)
\nonumber \hbox{Identities } A_\alpha: && \frac{2}{\sqrt{g}} \sum \frac{\partial \frak{K}^\sigma_\tau}{\partial w^\sigma} + \sum K_{\mu\nu} g^{\mu\nu}_\tau \equiv 0,
\\
\hbox{Identities } A_\beta: && \frac2{\sqrt{g}} \sum \frac{\partial (\frak{K}^\sigma_\tau + \frak{U}^\sigma_\tau)}{\partial w^\sigma} \equiv 0, \qquad\qquad \hbox{for}\; \tau = 1, 2, 3, 4
\\
\nonumber \hbox{Identities } A_\gamma: &&  \frac{2}{\sqrt{g}} \sum \frac{\partial(\frak{K}^\sigma_\tau + \frak{U}^{*\sigma}_\tau)}{\partial w^\sigma} \equiv 0.
\end{eqnarray}
In this case, I have always assumed the terms with the $\frak{K}^\sigma_\tau$ to be first, which appears practical; what is more, I added constant factors so that the left sides are always the same four vector components.

\bigskip

\centerline{\S 6.}
\medskip

\centerline{\textbf{Transition to conservation laws}.}
\bigskip

What I am to say, in the course of the systematic train of thought, on the special construction of the invariant $K$, which is the basis of modern gravitational theory, is so close to Einstein's pertinent investigations that I prefer to postpone it until the following paragraph, and here I would like to follow the fundamental transition to the differential laws of the conservation of momentum and energy, and  survey  the various forms in which these laws have appeared in the literature. For the material field with which we are concerned, the ten gravitational equations, as I have already pointed out in the introduction under (3), are particularly simple in our description. I will put German here instead of the Latin letters and then have
\begin{equation}%(38)
\frak{K}_{\mu\nu} - \chi \frak{T}_{\mu\nu} = 0.
\end{equation}
Instead, of course, I can also write the 16 equations:
\begin{equation}%(39)
\frak{K}^\sigma_\tau - \chi \frak{T}^\sigma_\tau = 0.
\end{equation}
\textit{All we have to do now is that we have to insert the values following from this of the $\frak{K}_{\mu\nu}$ respectively the $\frak{K}^\sigma_\tau$ in the identities set up for the invariant $K$.} The matter is so simple that I can compile and explain the results in a tabular form.

\newpage

\noindent XXXII. Differential Form of Conservation Laws in Gravitational Theory. 579

\bigskip

I start with the identities $A_\alpha$ to $A_\gamma$ (37):

1. From $A_\alpha$, we have following division with $\frac{2\chi}{\sqrt{g}}$:
\begin{equation}%(40)
\sum\frac{\partial\frak{T}^\sigma_\tau }{\partial w^\sigma}+\frac12\sum \frak{T}_{\mu\nu}g^{\mu\nu}_\tau = 0.
\end{equation}
 These are the conservation laws for the energy components of the material field as such, as they are found everywhere in the literature.

2. Of course, I can also write what is certainly new:
\begin{equation}%(41)
\sum \frac{\partial \frak{T}^\sigma_\tau}{\partial w^\sigma} + \frac1{2 \chi} \sum \frak{K}_{\mu\nu} g^{\mu\nu}_\tau = 0.
\end{equation}

3. Hereby it is wholly equivalent if I conclude from the $A_\beta$:
\begin{equation}%(42)
\sum \frac{\partial \left( \frak{T}^\sigma_\tau + \frac1{\chi} \frak{U}^\sigma_\tau \right)}{\partial w^\sigma} = 0.
\end{equation}

These are, in essence, the conservation laws as set out by Lorentz in Part III of his series of articles, cited above p.~482, formula (79). (The direct identification is only somewhat qualitative inasmuch as Lorentz did not arrange the $\delta I_1$ according to the $\delta g^{\mu\nu}$, but the $\delta g_{\mu\nu}$, however it cannot be doubted, because he is able to derive the same infinitesimal transformation $\delta w^\tau = p^\tau$ (with constant $p^\tau$), which has led us to the identities $A_\beta$.%
\footnote{[In fact, Mr Vermeil has confirmed the identity of the results of the two expressions by direct calculation. K.]}

4. Finally, the same relations are also written according to $A_\gamma$:
\begin{equation}%(43)
\sum \frac{\partial \left( \frak{T}^\sigma_\tau + \frac1{\chi} \frak{U}^{*\sigma}_\tau \right)}{\partial w^\sigma} = 0.
\end{equation}
The $K^*$ (formula (21)) and thus the $\frak{U}^{*\sigma}_\tau$ are only to be appropriately particularized in order to obtain the well-known Einstein formulas:
\begin{equation}%(44)
\sum \frac{\partial \left( \frak{T}^\sigma_\tau + \frak{t}^{\sigma}_\tau \right)}{\partial w^\sigma} = 0.
\end{equation}
Further details will be given in the following paragraph. In any case, it is already understood here that the left sides of the Einstein relations, multiplied by $\sqrt{g}$, are vector components as well as the exactly agreeing left-hand sides of (41), (42). I emphasize this only because the situation does not seem to be clearly recognized everywhere.

\bigskip

37 *

\newpage

580 On the Erlangen Program.
\bigskip

We now return to the original summary (31), (31$'$) of the identities $B$:
$$
\left[ \frac1{\sqrt{g}} \sum \frac{\partial \sqrt{g} \varepsilon^\sigma}{\partial w^\sigma} = \frac1{\sqrt{g}} \sum
\frac{\partial \sqrt{g} (\eta^\sigma + 2 \sum K^\sigma_\tau p^\tau)}{\partial w^\sigma} \right] \equiv 0.
$$
Here, for $K^\sigma_\tau$, we write the value $\chi T^\sigma_\tau$ which follows from the gravitational equations of the field, a new vector, which may be called $e^\sigma$, replaces the vector $\varepsilon^\sigma$:
\begin{equation}%(45)
e^\sigma = \eta^\sigma + 2 \chi \sum T^\sigma_\tau p^\tau.
\end{equation}
\textit{This new vector, as I assert, is exactly what Hilbert has designated as the energy vector in his note (with the restriction to the electromagnetic case)} (so that Hilbert's conservation theorems are summarized in the one equation)
\begin{equation}%(46)
\frac1{\sqrt{g}} \sum \frac{\partial \sqrt{g} e^\sigma}{\partial w^\sigma}.
\end{equation}

For the proof I remark:

(a) As regards the part which originates from the ``matter'' in (45), i.e., the term $2\chi T^\sigma_\tau p^\tau$, this is true, if I first set $\chi$ to 1 in accord with Hilbert, after a reasonable change of the notation with the assumptions which Hilbert makes in formula (19) of his note and in the subsequent sentences, without further ado.

(b) Then, as far as the ``gravitational part'' is concerned, Herr Freedericks has long ago given me the terms which appear at first unclear, as Hilbert loc.~cit.\ in formulas (8), (9), and (14) , computationally combined, and has come exactly to the expression which I have introduced as $\eta^\sigma$ in (11).%
\footnote{[Hilbert apparently chose the appearance of $e^\sigma$, which seems very complicated at first, in order to allow the vector character of this quantity to stand out from the outset. K.]}

Now the formula (46), even if I divide the factor $\frac{2 \chi}{\sqrt{g}}$, looks quite different from the formulas (42), (43).
The subject matter is quite clear, however, if I expand (46) according to the scheme (33) to (36) into 4 + 16 + 40 + 80 equations:

The first four equations are as follows:
\begin{equation}%(47)
\sum \left( \frak{U}^\sigma_{\tau,\sigma} + \chi \frak{T}^\sigma_{\tau,\sigma} \right) = 0,
\end{equation}
and so exactly match the equations (42).

The following 16 equations will be:
\begin{equation}%(48)
\frak{U}^\sigma_\tau + \chi \frak{T}^\sigma_\tau + \sum_{\sigma'} \frak{U}^{\sigma'\!,\sigma}_{\tau,\sigma'} = 0.
\end{equation}

\newpage

\noindent XXXII. Differential Form of the Conservation Laws in Gravitational Theory. 581
\bigskip

\noindent This is only a special notation of the field equations (39), since
$\frak{U}^\sigma_\tau + \sum_{\sigma'} \frak{U}^{\sigma'\!,\sigma}_{\tau,\sigma'}$
is identical to $-\frak{K}^\sigma_\tau$ in (34).

The remaining 40 + 80 equations are, however, in agreement with the identities (35), (36); they have nothing to do with the material field we are looking at.

In essence, the Hilbert statement (46) is thus reduced to the conservation laws (42); what is added are anyway known equations. On the other hand, the proposition has the advantage that it not only simply asserts itself somewhat invariant-theoretically, but also that the quantity $e^\sigma$ occurring in it can be briefly characterized in invariant-theoretical terms: \textit{it contains the helping-vector $p^\tau$, but otherwise the $T_{\mu\nu}$, the $K_{\mu\nu}$ and therefore is a differential-quotient depending on a co-gredient vector}.

With the explicit statements made in this section on the various forms of the conservation statements, as one can see, will be completed that which was expressed only in an undefined way in the numbers (6) to (8) of my previous note.

\bigskip

\centerline{\S 7.}
\medskip

\centerline{\textbf{More details on Einstein's formulation of conservation laws}.}
\bigskip

I now have to add how the quantity I designated by $K^*$ has to be particularized in order to arrive at Einstein's final formula:
$$%(44)
(44) \hspace{4.5cm} \sum \frac{\partial (\frak{T}^\sigma_\tau + \frak{t}^\sigma_\tau)}{\partial w^\sigma} = 0, \hspace{4.5cm}
$$
also to say a few things about the simplification achieved with it.

I like to refer to Einstein's above-mentioned description in the meeting reports of the Berlin Academy of October 1916. Einstein assumes that the invariant $K$ (which he calls $G$) contains the second differential quotients of the $g^{\mu\nu}$ only linearly, multiplied by the functions of $g^{\mu\nu}$ itself. It is therefore possible to eliminate said differential quotients from the integral $I = \int\!\int\!\int\!\int K d\omega$ by partial integration, i.e.,
\begin{equation}%(49)
K = G^* + {\rm Div}
\end{equation}
where $G^*$ is a function of only the first differential quotients. In particular, Einstein gives $G^*$ the value:
\begin{equation}%(50)
G^* = \sum_{\mu\nu\varrho\sigma} g^{\mu\nu} \{ \Gamma^\sigma_{\mu\varrho} \Gamma^\varrho_{\nu\sigma} - \Gamma^\varrho_{\mu\nu} \Gamma^\sigma_{\varrho\sigma} \},
\footnote{[On reprinting, the sign of $G^*$ and $\frak{G}^*$ was here and hereafter amended, according to the circumstances not sufficiently taken into account in the first publication, that the sign of $ds^2$ is taken, according to Einstein, as in the footnote, page 569 of this treatise; then Einstein's $G$ is identical with Hilbert's $K$. K.]}
\end{equation}
\newpage

582 On the Erlangen Program.

\bigskip

\noindent and $-\Gamma^\varrho_{\mu\nu}$ the so-called symbols of the second kind are understood as
\begin{equation}%(51)
-\Gamma^\varrho_{\mu\nu} = \sum_\tau \frac{g^{\varrho\tau}}{2} \left( \frac{\partial g_{\mu\tau}}{\partial w^\nu} + \frac{\partial g_{\nu\tau}}{\partial w^\mu} - \frac{\partial g_{\mu\nu}}{\partial w^\tau} \right).
\footnote{See the implementation of the interpolation on pages 110, 191 of Weyl's book.}
\end{equation}
Obviously, this $G^*$ is invariant under affine transformations of the $w$.

The further Einsteinian concluding formulas now follow immediately from our earlier approaches if we set
\begin{equation}%(52)
K^* = G^*, \quad \hbox{thus} \quad \frak{K}^* = \frak{G}^*;
\end{equation}
we only have to take
$$ \chi = 1 $$
afterwards to have full agreement.

There are actually only two points:

a) According to (21), we have
\begin{equation}%(53)
\frak{K}_{\mu\nu} \equiv \frak{G}^*_{\mu\nu}.
\end{equation}
But $\frak{G}^*_{\mu\nu}$ is formally simpler than the $\frak{K}_{\mu\nu}$, because $\frak{G}^*$ contains only the first order differential quotients of $g^{\mu\nu}$:
\begin{equation}%(54)
\frak{G}^*_{\mu\nu} = \frac{\partial \frak{G}^*}{\partial g^{\mu\nu}} - \sum \frac{\partial \left( \frac{\partial{{\frak{G}}{}^{\ast}}}{\partial g^{\mu\nu}_\varrho} \right)}{\partial w^\varrho}.
\end{equation}
As such, the $\frak{G}_{\mu\nu}^*$ of Einstein occur in fact instead of the $\frak{K}_{\mu\nu}$ in the field equations (formula (7) of his article). It may be said that by the introduction of the $\frak{G}_{\mu\nu}^*$, a special property of the $\frak{K}_{\mu\nu}$, namely not to contain differential quotients of the $g^{\mu\nu}$ of higher than the second order, has been visibly revealed.

b) Furthermore, for the $\frak{U}^{*\sigma}_\tau$ according to (16), (22) we have the simple formulas
\begin{equation}%(55)
\frak{U}^{*\sigma}_\tau = \frac12 \left( \frak{G}^* \delta^\sigma_\tau - \sum \frac{\partial \frak{G}^*}{\partial g^{\mu\nu}_\sigma} g^{\mu\nu}_\tau \right).
\end{equation}
These $\frak{U}^{*\sigma}_\tau$ are actually abridgements of the general $\frak{U}^\sigma_\tau$, but the result of the divergence formation
$$
\sum \frac{\partial \frak{U}^{*\sigma}_\tau}{\partial w^\sigma}
$$
is again the same. Hence the reduction of the formulas brings only the simplification into clear view, which is the result of the $K$-type of construction which is relevant to us.

\newpage

\noindent XXXII. Differential Form of Conservation Laws in Gravitational Theory. 583

\bigskip

\textit{The $\frak{U}^{*\sigma}_\tau$, defined by (55), divided by $\chi$, are now directly Einstein's $\frak{t}^\sigma_\tau$}:
\begin{equation}%(56)
\frac1{\chi} \frak{U}^{*\sigma}_\tau = \frak{t}^\sigma_\tau.
\end{equation}
In fact, in formula (20) of his treatise --- based on a completely different calculation ---, by taking $\chi = 1$, Einstein gives exactly the values on the right hand side of (55) for his $\frak{t}^\sigma_\tau$.

Let us take the $\frak{t}^\sigma_\tau$ for the $\frac1{\chi} \frak{U}^{*\sigma}_\tau$ in (43), we obtain the equations (44), which was to be proved.

\bigskip

\hfil---------------------------\hfil

\bigskip

I would like to add a small addition to these developments. In his ``Cosmological Reflections on the General Relativity Theory'',%
\footnote{Meeting reports of the Berlin Academy of February 8, 1917.}
Einstein has, as we know, proposed to modify the fundamental field equations of gravitation to the effect that --- in our notation --- instead of (3)
\begin{equation}%(57)
K_{\mu\nu} - \lambda g_{\mu\nu} - \chi T_{\mu\nu} = 0,
\end{equation}
where $\lambda$ is a constant. Since
$$
\frac{\partial \sqrt{g}}{\partial g^{\mu\nu}}:\sqrt{g} = - \frac12 g_{\mu\nu}
$$
so we can also write (57) as
\begin{equation}%(58)
\bar K_{\mu\nu} - \chi T_{\mu\nu} = 0, \quad \hbox{or also} \quad {\bar K}{}^\sigma_\tau - \chi T^\sigma_\tau = 0,
\end{equation}
where
\begin{equation}%(59)
\bar K = K + 2 \lambda.
\end{equation}

\textit{Now for this $\bar K$ all the conditions on which the identities for $K$ have been set up in paragraphs 2 to 5 apply}. Thus, for the $\bar K$, we may, for example, write the identities (37) in which we have to replace the $U^\sigma_\tau$ only by  $\bar U^\sigma_\tau$, where according to (16) will become
\begin{equation}%(60)
\bar U^\sigma_\tau = U^\sigma_\tau + \lambda \delta^\sigma_\tau.
\end{equation}
We thus obtain conservation laws, as in the past, comparable to the formula (42)
\begin{equation}%(61)
\sum \frac{\partial \left( \frak{T}^\sigma_\tau + \frac{1}{\chi} \bar{\frak{U}}{}^\sigma_\tau \right)}{\partial w^\sigma} = 0,
\end{equation}
where we may now modify the $\bar{\frak{U}}{}^\sigma_\tau$ by substituting for $\bar K$
\begin{equation}%(62)
\bar K^* = \bar K + \frak{Div}.
\end{equation}

\newpage

584 On the Erlangen Program.
\bigskip

\noindent Specifically, for ${\bar K}^*$, according to our latest developments we will take
\begin{equation}%(63)
\bar G^* = G^* + 2 \lambda, \quad \hbox{i.e.} \quad \bar{\frak{G}}{}^* = \frak{G}^* + 2 \lambda \sqrt{g}.
\end{equation}
If we then write from (55), (56):
\begin{equation}%(64)
\bar{\frak{t}}{}^{\sigma}_\tau = \frac1{2\chi} \left( \bar{\frak{G}}{}^*\delta^\sigma_\tau - \sum \frac{\partial \bar{\frak{G}}{}^*}{\partial g^{\mu\nu}_\sigma} g^{\mu\nu}_\tau, \right).
\end{equation}
we will now have
\begin{equation}%(65)
\sum \frac{\partial (\frak{T}^\sigma_\tau + \bar{\frak{t}}{}^\sigma_\tau)}{\partial w^\sigma} = 0.
\end{equation}
This corresponds to the statement made by Einstein in his latest publication.%
\footnote{Conference reports of the Berlin Academy of May 16, 1918, p. 456.}

\bigskip

\hfil------------------------\hfil

\bigskip

\centerline{\S 8.}
\medskip

\centerline{\textbf{Conclusion}.}
\bigskip

The relations which the developments so far given to the works of Einstein, Hilbert and Lorentz and Weyl quoted by me are, in detail, even closer, than by the mere comparison of the results obtained. Many of the formulas that occur in the intermediate reflections are also found there, but not in the uniform formulation which I have observed. It is very interesting to follow this in detail. The closest to my developments are those of Lorentz, which, however, are soon confined to such infinitesimal transformations $\delta w^\tau = p^\tau$, whose $p^\tau$ are independent of the $w$. Einstein regards such $p^\tau$ as corresponding to the affine transformations of $w$, Weyl (as I have in my previous note) those $p^\tau$, which are otherwise arbitrary, but disappear in a suitable manner on the boundary of the domain of integration.%
\footnote{Thus, already in a essay ``On Gravitation Theory'' (in Vol. 54 of the Annals of Physics), which was completed before my note, but was only published after it.}

I must also not omit to thank Miss~N\"other for encouraging participation in my new work, where the mathematical ideas which I used in the adaptation to the physical question for the integral $I_1$ have in general been worked out, and will in the near future in these news be published.\footnote{[On the 26th of July I published the main records of Miss~N\"other of the Society of Sciences. The note itself has also been published in the G\"ottingen News, 1918, pp. 235--257, under the title ``Invariant Variational Problems.'']---}

\newpage

\noindent XXXII. Differential Form of Conservation Laws in Gravitational Theory. 585

\bigskip

%%% ``principle theorem'', ``main theorem'', ``fundamental theorem''
[The ``fundamental theorem'' set out in \S 2 above is a special case of the following extensive theorem proved by Fr.~N\"other in the place indicated:

``If an integral $I$ is invariant with respect to a $G_\varrho$ (that is, a continuous group with $\varrho$ essential parameters), then $\varrho$ linearly independent combinations of Lagrange's expressions become divergences''

However, as regards Hilbert's assertion contained in XXXI (see pp.~561 and 565 of this edition), the exact formulation according to Miss~N\"other is the following:

``If an integral $I$ allows the translation group, then the energy relations become improper if and only if $I$ is invariant under an infinite group containing the translation group as a subgroup.''

Moreover, as regards the theorem of Hilbert and of XXXI: it also holds that there are four relations between the field equations of the theory of relativity,  by Miss~N\"other this is generalized. Her theorem is as follows: ``If the integral $I$ is invariant under a group with $\varrho$ arbitrary functions in which these functions occur up to the $\sigma$-th derivative, then there are $\varrho$ identical relations between the Lagrangian expressions and their derivatives to the $\sigma$-th order.'' K.]

\end{document}